\documentclass[12pt]{article}
\newcommand{\beq}{\begin{equation}}
\newcommand{\eeq}{\end{equation}}
\newcommand{\bea}{\begin{eqnarray}}
\newcommand{\eea}{\end{eqnarray}}
\newcommand{\du}{d_{\cal U}}

\newcommand{\lagr}{{\cal L}}
\newcommand{\opr}{{\cal O}}
\newcommand{\unp}{{\cal U}}
\newcommand{\bra}{\langle}
\newcommand{\ket}{\rangle}
\newcommand{\Mbh}{M_{\rm BH}}
\newcommand{\Mpl}{M_{\rm D}}

\usepackage{graphicx}
 \usepackage{setspace}
 \usepackage{amsfonts}
 \usepackage{amsmath}  
\begin{document}
\pagestyle{empty}
\begin{center}
\vskip 2cm
{\bf \LARGE Differentiating Unparticles from Extra Dimensions Via Mini Black Hole Thermodynamics}
\vskip 1cm
J. R. Mureika \\

{\small \it Department of Physics, Loyola Marymount University, Los Angeles, CA  90045-2659} \\
Email: jmureika@lmu.edu
\end{center}

{\noindent{\bf Abstract} \\
A thermodynamics-based method is presented for differentiating mini black hole creation mechanisms in high energy parton collisions, including scenarios with large compactified extra dimensions and unparticle-enhanced gravity with real scaling dimension $\du$.   Tensor unparticle interactions are shown to mimic the physics of  $(2\du-2)$ non-integer extra spatial dimensions.  This yields unique model-dependent production rates, Hawking temperature profiles, and decay multiplicities for black holes of mass $\Mbh \sim 1-15~$TeV that may be created at the LHC and other future colliders.
}


\section{Introduction}
On the eve of the LHC's historic run, anticipation is high for the discovery of new physics.  Of the many plausible scenarios that could unfold, one of the most intriguing is confirmation of the existence of compactified extra spatial dimensions as posited by Arkani-Hamed, Dimopoulos, and Dvali (ADD) \cite{add}.  This model addresses the hierarchy problem by suggesting that the scale of quantum gravity is not defined by the traditional Planck mass ($\sqrt{G}^{-1} = M_{\rm Pl} \sim 10^{16}~$TeV), but is instead described by a fundamental scale $\Mpl \sim 1~$TeV coincident with that of electroweak unification.   Gravitational couplings are thus much stronger at short distance scales, which can induce the creation of mini black holes in high energy collisions \cite{fischler,landsberg,giddings}.  

The introduction of unparticle physics in 2007 \cite{georgi,cheung1} has added yet another framework ripe for TeV-scale testing,  and has precipitated a flurry of new phenomenological and modifications to fundamental theory.    Unparticle stuff is an otherwise invisible high energy conformally-invariant field with non-intuitive phase space structure that may become accessible at the LHC or in other future colliders.  The theory is predicated on the existence of a weakly-coupled Banks-Zaks (BZ) field \cite{bz} exchanging with the SM a massive particle $M_{\unp}$  via suppressed non-renormalizable interactions of the form $\lagr = \frac{1}{M^k_{\unp}} \opr_{SM}\opr_{BZ}$, where the $\opr$ represent the respective field operators of dimensions $d_{SM}$ and $d_{BZ}$ respectively, and $k = d_{SM} + d_{BZ} - 4$. 

Below some energy scale $\Lambda_\unp < M_\unp$, the coupling begins to run as the BZ field undergoes dimensional transmutation to become ``unparticle stuff'', represented by the operator 
\beq
\opr_{BZ} \rightarrow C_\unp \lambda^{d_{BZ} - d_\unp}\opr_\unp
\eeq
of dimension $d_\unp \ne d_{BZ}$.  The new interaction picture is given as 
\beq
\lagr = \frac{\kappa}{\Lambda^{k_\unp}} \opr_{SM} \opr_\unp~,~\kappa = C_\unp \left(\frac{\Lambda_\unp}{M_{BZ}}\right)^k~~,
\eeq 
with $k_\unp = d_{SM}+d_\unp-4$.   Postulating that $\Lambda_\unp \sim 1~$TeV provides a wealth of new physics that can not only be observabled at the LHC, but also stands to modify astrophysical and cosmological mechanisms.

The matrix element 
\beq
\bra 0 | \opr_\unp(x) \opr^\dagger_\unp(0)|0\ket  = \int \frac{d^4p}{(4\pi)^4} e^{iPx} |\bra 0 | \opr_\unp(0)|P\ket|^2 \rho(P^2)
\eeq
for an unparticle of four-momentum $P$ constrains the spectral density function to be of the form
\beq
|\bra 0 | \opr_\unp(0)|P\ket|^2 \rho(P^2) = {A_d}_{\cal U} \theta(P^0) \theta(P^2) (P^2)^{\du -2}~,
\eeq
with the coefficient ${A_d}_{\cal U} = \frac{16 \pi^{5/2}}{(2\pi)^{2n}} \frac{\Gamma(n+1/2)}{\Gamma(n-1)\Gamma(2n)}$.  Comparing this expression to the standard phase space $A_n \theta(P^0) \theta(P^2) (P^2)^{n -2}$ of $n$ interacting particles of total momentum $P$ results in the interpretation that unparticle stuff looks like a collection of non-integer ($\du$) indivisible particles \cite{georgi}.  Several explanations of the physical nature of unparticle stuff include a composite Banks-Zaks particle with a continuum of masses \cite{kraz,nikolic,mcdonald2}, or alternatively a Sommerfield-like model of massless fermions coupled to a massive vector field \cite{georgi2}.

\section{Unparticle-enhanced gravity}
The literature on unparticle phenomenology is vast and cannot be completely cited herein.  This letter focuses on the application of unparticle physics to the gravitational sector \cite{goldberg,hsu1,damora,jrmplb}.  In the current model, it is assumed that a tensor unparticle field couples simply to matter via the stress energy tensor  \cite{goldberg}
\beq
T^{\mu\nu} + T_\unp^{\mu \nu}~~,~~ T_\unp^{\mu \nu} \sim  \sqrt{|g|}T^{\alpha \beta}\opr^\unp_{\alpha \beta} \; g_{\mu \nu}~~.
\eeq
Other possible couplings such as $T^{\mu \alpha} O_{\alpha}^{\nu}$ are possible, but will not be dealt with here.  The term ${\cal T}_\unp^{\mu \nu}$ resembles an effective cosmological constant-like influence, which can be geometrized in the spacetime metric and thus contribute to modifications of general relativistic effects such as black hole formation ({\it e.g.} similarly to the Schwarzschild-deSitter metric \cite{rindler}).   It is also possible that such a term can also co-exist with a regular cosmological constant term and introduce hidden-sector supersymmetry-like cancelations that could address the cosmological constant problem. 

The non-relativistic potential corresponding to the unparticle coupling is \cite{goldberg,damora}
\beq
V(r) = -\frac{2Gm_1 m_2}{r} \left[1 + \frac{2}{\pi^{2\du-1}}\; \frac{\Gamma(\du + \frac{1}{2})\Gamma(\du -\frac{1}{2})}{\Gamma(2\du)} \left(\frac{R_*}{r}\right)^{2\du-2}\right] \equiv V_N(r) \left[1+\Gamma_{\du}\left(\frac{R_*}{r}\right)^{2\du-2}\right] ~ ,
\label{tensorV}
\eeq
where the effective unparticle interaction scale  is \cite{goldberg}
\beq
R_* = \Lambda_\unp^{-1} \left(\frac{M_{Pl}}{\Lambda_\unp}\right)^\frac{1}{\du-1} \left(\frac{\Lambda_\unp}{M_\unp}\right)^\frac{d_{BZ}}{\du-1}~~.
\eeq 

A suitable approximation of the temporal and radial unparticle-enhanced metric coefficients is \cite{jrmplb}
\beq
g_{00} \approx  1-\frac{2Gm}{r}\left(1+ \Gamma_{\du} \left(\frac{R_*}{r}\right)^{2\du-2}\right)~~,~~g_{11} = -g_{00}^{-1}
\label{unpmetric}
\eeq
The choice of this form ensures that the proper Newtonian potential (\ref{tensorV}) is recovered in the weak-field limit.   Additionally, for $r \ll R_*$ the potential will resemble that of a $(4+n)$-dimensional spacetime, where in the case of unparticles $n = 2\du-2$ \cite{jrmplb},
\beq
\Phi(r) \sim \frac{G\Mbh\Gamma_{\du}}{r} \left(\frac{R_*}{r}\right)^{2\du-2}~~.
\eeq
This dimensional correspondence between the ADD model and unparticle physics was previously noted in \cite{cheung2}, vis-a-vis the momentum power signature of the phase space.  It is thus the case that tensor unparticle interactions will mimic the effects of compactified extra dimensions, and thus for a parton collision of energy $\Mbh$ will induce an $\Mpl$-independent horizon distance  \cite{jrmplb}
\beq
r_{H_\unp} = \left[ 2\Mbh\Gamma_{\du}\Lambda_\unp^{-1} \left(\frac{M_\unp}{\Lambda_\unp}\right)^{-2d_{BZ}}\right]^\frac{1}{2\du-1} \Lambda_\unp^{-1} ~~.
\label{unphorizon}
\eeq
This can be compared to the standard ADD horizon distance,
\beq
r_{ED}= \frac{1}{\Mpl \sqrt{\pi}} \left(\frac{\Mbh}{\Mpl} \cdot \frac{8\Gamma\left(\frac{n+3}{2}\right)}{n+2}\right)^\frac{1}{n+1}~~.
\label{addhorizon}
\eeq
Note that the Planck mass $M_{\rm Pl}$ drops out of the expression for $r_\unp$, allowing the unparticle gravity effects to become manifest at the appropriate energy scales $\Lambda_\unp$.  

\begin{figure}[h]
\begin{center}
\leavevmode
\includegraphics[scale=0.6]{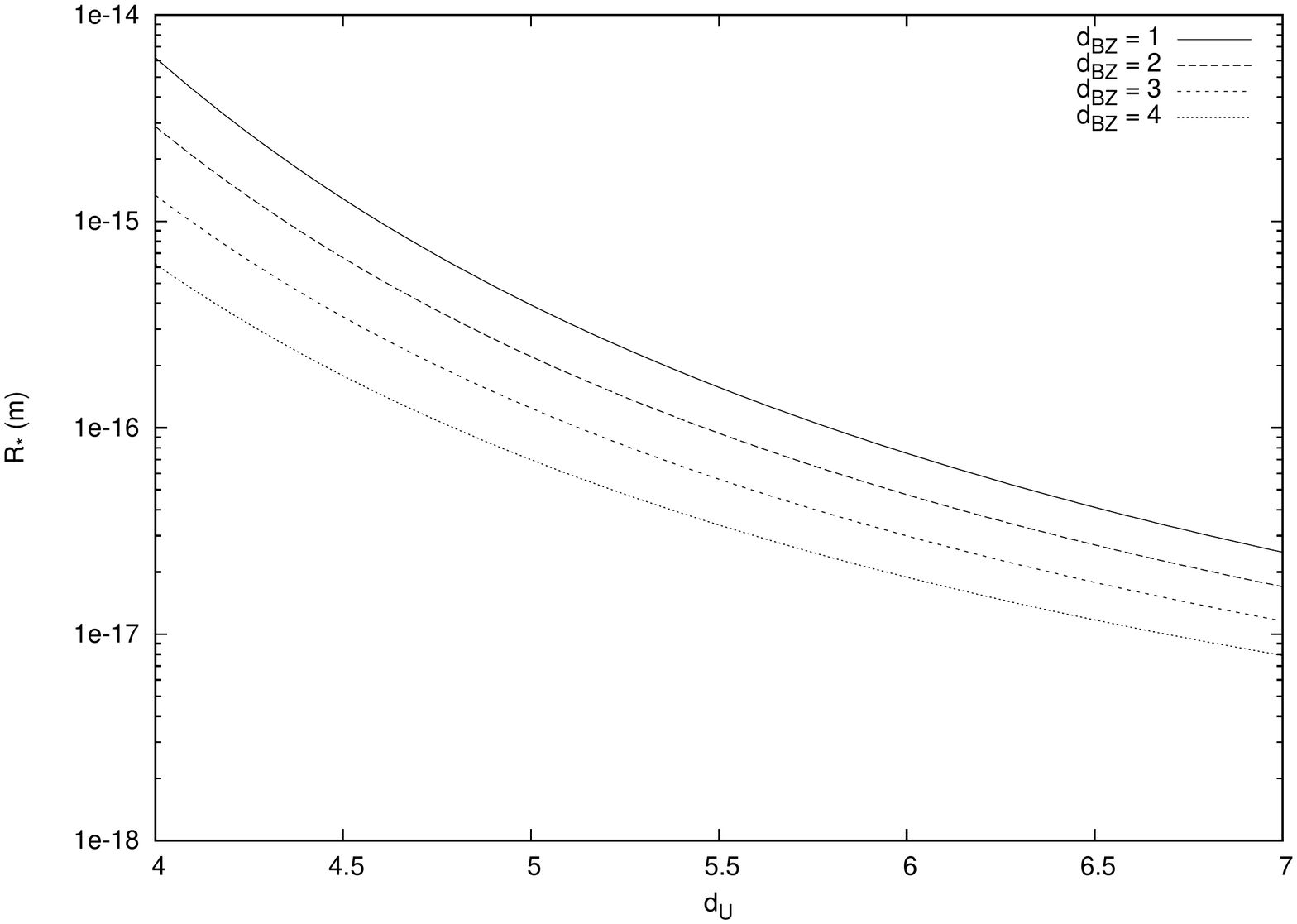}
\caption{Unparticle interaction scale $R_*$ (in metres) as a function of dimension $d_\unp$ for $d_{BZ} = 1-4$, with $\Lambda_\unp = 1~$TeV and $M_\unp = 10~$TeV.}
\label{fig1}
\end{center}
\end{figure}

To ensure that unitarity is not violated for tensor unparticles, the condition $d_\unp \geq 4$ must be imposed \cite{nakayama,grinstein}.  For $d_{BZ}= 1-4$, the unparticle scale $R_*$ is on the order of $10^{-14}-10^{-15}~$m, with decreasing values for increasing $d_\unp$ (Figure~\ref{fig1}).  The sensitivity of the results to the Banks-Zaks dimension $d_{BZ}$ is generally small (less than an order of magnitude for a range of dimensions), so long as the BZ-messanger mass $M_\unp$ is also small ({\it i.e.} 10-100~TeV).    These values fall in an allowable parameter space for both astrophysical and gravitational unparticle phenomena \cite{goldberg,mcdonald1}.   For larger $M_\unp$ and $d_{BZ}$, the horizon radius shrinks and the black hole cross-section becomes insignificant for the LHC.  As $\du$ grows, however, the geometric cross-section $\sigma_\unp = \pi r_{H_\unp}^2$ increases \cite{jrmplb}.  Choosing $M_\unp = 10~$TeV and $\Lambda_\unp = 1~$TeV, the minimum geometric cross-section is thus between $\sigma_\unp \sim 0.1~$pb ($d_{BZ} = 4$) and $\sigma_\unp \sim 10~$pb ($d_{BZ} = 1$), with each approaching the range $\sigma_\unp \sim 1-10~$pb for increasing $\du$. 
  
The fact that black holes of a given mass will in general be smaller in the unparticle scenario than in the ADD mechanism is itself a first indication of model dependence, but there are other cross-checks that yield further discrepancies.  If such singular objects are observed at the LHC or future colliders, it would thus be instructive to determine which mechanism might play a role in their creation.\footnote{The following comparison is strictly between ADD black holes and unparticle-enhanced ones; it does not include Randall-Sundrum black holes \cite{chr1,rsbh}, or those posited in Reference~\cite{chr2}.}  

\begin{figure}[h]
\begin{center}
\leavevmode
\includegraphics[scale=0.6,angle=-90]{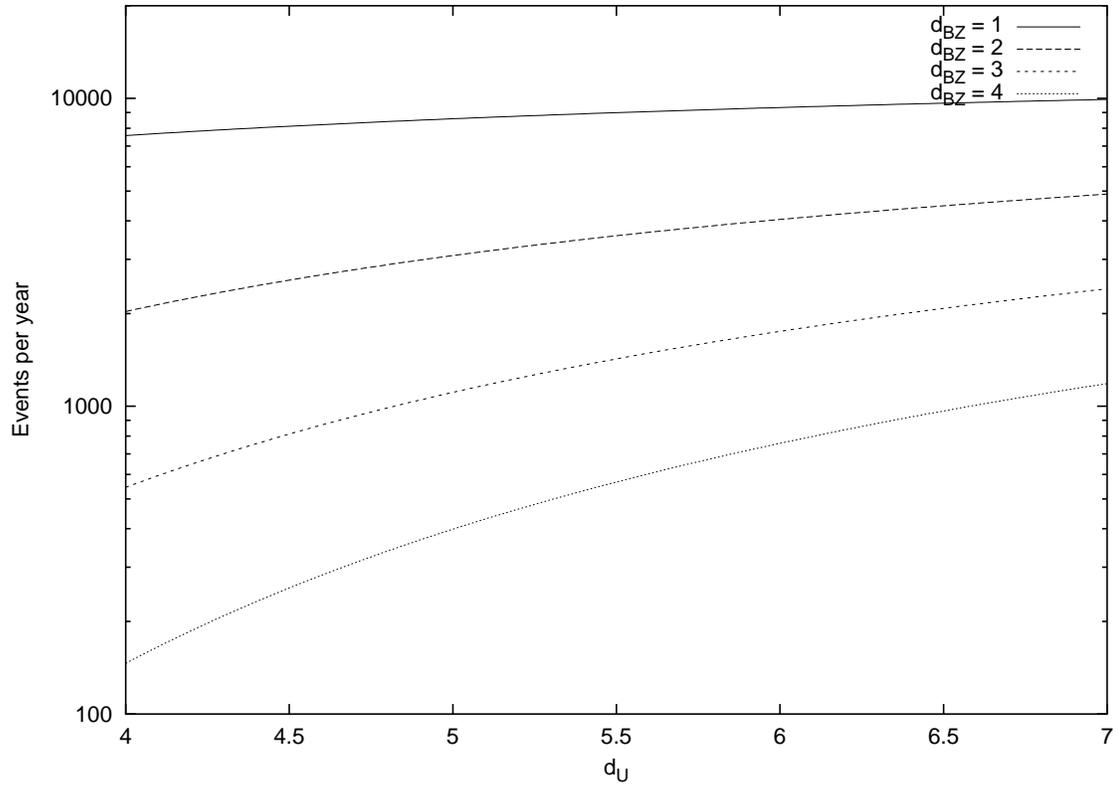}
\caption{Annual integrated production rate estimates for mini black holes generated in the unparticle-enhanced scenario, as a function of $d_\unp$ for $d_{BZ} = 1-4$, with $\Lambda_\unp = 1~$TeV, $M_\unp = 10~$TeV, and $\Mbh = 5~$TeV.}
\label{fig2}
\end{center}
\end{figure}

The integrated parton-parton cross-section can be obtained in the usual fashion as $\sigma_{pp\rightarrow BH}(\Mbh) = h(\frac{\Mbh}{\sqrt{s}})\sigma_\unp(\Mbh)$, where the function $\frac{\Mbh}{\sqrt{s}}$ is the integrated cross-section function evaluated at the ``ij''-parton-parton center-of-mass energy squared $s_{ij} \leq s$ (see \cite{giddings} for a expanded treatment).   Figure~\ref{fig2} demonstrates the annual integrated production rates for mini black holes of mass $\Mbh \geq 5~$TeV at the LHC, using the CTEQ5 parton distribution functions with $h(0.357) \sim 0.02$ and assuming a peak luminosity of 30~${\rm fb^{-1} yr^{-1}}$  \cite{landsberg, giddings}.  In the most liberal scenario of $\du = 1$, one can expect just under $10^4$ such black holes per year,  {\it i.e} a frequency of  1~${\rm hr}^{-1}$.  This is considerably less than the production rate for the ADD scenario, which predicts an optimal frequency of 1~Hz for a cross-section of 100~pb.  For increasing value of $\du$, the production rates for low-mass black holes will be reduced by over an order of magnitude.  As the collision energy (and hence $\Mbh$) increases, the production rates are further suppressed by the integrated cross-section.  Such production rates nevertheless suggest that the unparticle-enhanced black hole mechanism will produce sufficient signals for detectability.

\section{Unparticle v. ADD black hole thermodynamics}
Beyond these simple calculations, much can be learned from a thermodynamical analysis of the mini black holes.  A characteristic to which one may appeal is the lifetime of the black hole, and whether or not there are measurable framework-dependent differences.  Whether there is a bulk volume of extra dimensions or an unparticle-driven mechanism, it is expected that a mini black hole will radiate primarily in the usual four-dimensional ``brane'' (loosely termed in the case of $n=0$ extra dimensions), emitting the complete spectrum of standard model particles with power per unit area $p$ according to the Stefan-Boltzmann law $\frac{P}{A_{BH}} = p = \sigma T_H^4$.   Since $A_{BH} \sim r_H^2$, the total power will be $P \sim r_H^{-2}$, and the entire energy of the black hole will be radiated in a time $\tau_{BH} = \frac{\Mbh}{P}$ \cite{landsberg}.  In the case of compactified extra dimensions, substitution of the horizon radius \ref{addhorizon} with $\Mpl = 1~$TeV yields a lifetime of approximately $\tau \sim 10^{(n+3)/(n+1)-27}~$seconds.  By a similar argument, unparticle-enhanced black holes should have lifetimes on the order $\tau_\unp \sim 10^{(2\du+1)/(2\du-1)-27}~$seconds, which for all values of $\du$ is comparable to the extra-dimensional result (for all $n$).  Thus, decay times will yield no discernible information, at least to any measurable degree of accuracy.

There is, however, a very elegant method for differentiation based on the thermodynamic nature of the hypothetical mini black holes.   If the LHC is to become a true black-hole factory, then the multiple events that are generated will yield a wealth of information via their decay remnants.  From such information, it is theoretically  possible to reconstruct the Hawking temperature spectrum $T_H$ for the given black-hole mass $\Mbh > \Mpl$ from resultant particle jets, primarily photons and electrons.  This possibility was first raised in Reference \cite{landsberg} as a novel method of not only verifying black hole evaporation processes, but also of investigating the dimensionality of the bulk.

In four dimensional general relativity, a black hole of mass $\Mbh$ and Schwarzschild radius $r_S$ has an effective temperature $T_H =  \frac{1}{4\pi r_S} = \frac{1}{8\pi \Mbh}$.  The generalized expression in $(4+n)$-dimensions  is \cite{landsberg,myersperry}
\beq
T_{ED} = \frac{n+1}{4\pi r_{ED}}~~.
\label{thed}
\eeq
The thermal spectrum that results from a black hole of mass $\Mbh$ can be obtained by assuming the decay products obey Planck's law \cite{landsberg}, and can  easily be extracted from (\ref{thed}) as
\beq
\log\left(\frac{T_{\rm ED}}{\Mpl}\right) = -\left(\frac{1}{n+1}\right) \log \left(\frac{M_{BH}}{\Mpl}\right) + {\cal T}_{\rm ED}~.
\label{edplot}
\eeq
The constant term ${\cal T}_{\rm ED}$ is a dimensionless function of the geometry and topology of spacetime, leaving a log-linear relationship whose slope uniquely determines the number of extra dimensions \cite{landsberg}.   It is perhaps surprising that specific information about the size of the spacetime manifold may be extracted from such a seemingly trivial result.

Although originally derived from quantum field theory for a variety of metric solutions, these temperature expressions $T_H$ have been shown to be robustly obtained from semi-classical arguments, for example as a function of the metric coefficients  \cite{poisson,mannkerner}
\beq
T_H = \frac{\sqrt{f^\prime(r_H) g^\prime(r_H)}}{2\pi}~~,~~ds^2 =  f(r)\; dt^2 - \frac{dr^2}{g(r)} -r^2\; d\Omega^2
\label{htemp}
\eeq
Using the unparticle-enhanced metric approximation given by (\ref{unpmetric}) and evaluating (\ref{htemp}) at the radius (\ref{unphorizon}) yields the expected value of

\beq
T_\unp = \frac{2\du-1}{4\pi r_{H_\unp}}~~.
\label{unptemp}
\eeq
which has a spectrum analogous to (\ref{edplot}),
\beq
\log\left(\frac{T_\unp}{\Lambda_\unp}\right) = -\left(\frac{1}{2\du-1}\right) \log \left(\frac{M_{BH}}{\Lambda_\unp}\right) + {\cal T}_\unp~~,
\label{unpplot}
\eeq
where again the dimensionless constant ${\cal T}_\unp$ contains information relevant to the unparticle phase space.   Strictly-speaking, the differences in slopes of (\ref{edplot}) and (\ref{unpplot}) are largely sufficient to identify the underlying mechanism.   Unlike the extra-dimensional case where the slope is restricted to rational values $\left\{-\frac{1}{1+n}~,~n\in \mathbb{Z}\right\}$, for ungravity the relationship is now determined by any real slope $\left\{-\frac{1}{2\du-1}~,~\du \in \mathbb{R} \right\}$.

For $\du \geq 4$, the minimum number of extra dimensions required to produce ``equivalent'' phenomenology to unparticle couplings is $n \geq 6$. The unparticle-enhanced frameworks will yield any real slope between $-\frac{1}{7}$ and 0.  Slopes smaller than $-\frac{1}{7}$ will rule out the unparticle mechanism in favor of spaces with $n < 6$ extra dimensions.  Current empirical evidence points to a maximum compactification size of $R \leq 100~$nm, which for a modified Planck mass of 1~TeV corresponds to $n \geq 3$.

The slopes will be equal whenever $\du = \frac{n+2}{2}$, in which case differentiation may not be possible with this information.  One can subsequently turn to the value of the constants ${\cal T}_{\rm ED}$ and ${\cal T}_\unp$.  Setting $\Mpl = \Lambda_\unp = 1~$TeV, the intercepts simplify to 
\beq
{\cal T}_{\rm ED} = \log\left(\frac{n+1}{4\sqrt{\pi}}\right) -\frac{1}{n+1} \log\left[\frac{8\Gamma\left(\frac{n+3}{2}\right)}{n+2}\right]~~,
\label{intercept1}
\eeq
and
\beq
{\cal T}_\unp = \log\left(\frac{2\du-1}{4\pi}\right) - \left(\frac{1}{2\du-1}\right) \log 2\Gamma_{\du} + \left( \frac{2d_{BZ}}{2\du-1} \right) \log \left(\frac{M_\unp}{\Lambda_\unp}\right)~~.
\label{intercept2}
\eeq
These two expressions share common terms, except for the appearance of the extra parameters $M_\unp$ and $d_{BZ}$ in the latter.  

Setting $\du=\frac{n+2}{2}$, a healthy exercise in algerabic manipulation reduces (\ref{intercept2}) to the form ${\cal T}_\unp = {\cal T_{\rm ED}} + \delta_{\cal T}$, where
\beq
\delta_{\cal T} = - \left(\frac{1}{n+1}\right) \log \left[\frac{n+2}{n+1}\cdot \frac{1}{\Gamma\left(1+\frac{n}{2}\right)}\right] + \left( \frac{2d_{BZ}}{n+1} \right) \log \left(\frac{M_\unp}{\Lambda_\unp}\right)+\log\left[2\sqrt{\pi}^\frac{n}{n+1}\right]~~.
\eeq
This result can also be obtained by manipulating the form of (\ref{unptemp}) to be of the form
$T_\unp = \xi T_{\rm ED}$ (with $\delta_{\cal T} = \log\xi$), where $\xi =  \left[2^{(n+1)} \sqrt{\pi}^{n}  \left(\frac{n+1}{n+2}\right) \Gamma\left(1+\frac{n}{2}\right) \left(\frac{M_\unp}{\Lambda_\unp}\right)^{2d_{BZ}}\right]^{\frac{1}{n+1}}$.  This is further confirmation that unparticle-enhanced black holes of mass $\Mbh$ will be of different size than those in ADD gravity.

At the unitarity threshold value ($\du = 4$) and assuming the most optimistic case of $d_{BZ} = 1, M_\unp \sim 10~$TeV, one finds $\delta_{\cal T} \sim 1$.  This difference grows for increasing $d_{BZ}$ and $M_\unp$, and decreases for increasing $n$, vanishing only at the critical value
\beq
M^c_\unp =\left[\frac{n+2}{n+1} \cdot \frac{2^{-(n+1)}\sqrt{\pi}^{-n}}{\Gamma\left(1+\frac{n}{2}\right)}\right]^{1/2d_{BZ}}~\Lambda_\unp~.
\eeq
It is straightforward to show that $M^c_\unp \ll 1~$TeV for all reasonable values of $d_{BZ}$ and $n$.  Since the unparticle mechanism requires the hierarchy $M_\unp > \Lambda_\unp$, the difference $\delta_{\cal T}$ effectively never vanishes.  The two mechanisms can thus always be differentiated, even in the case that the temperature spectrum slopes are equal.  

\begin{figure}[h]
\begin{center}
\leavevmode
\includegraphics[scale=0.6]{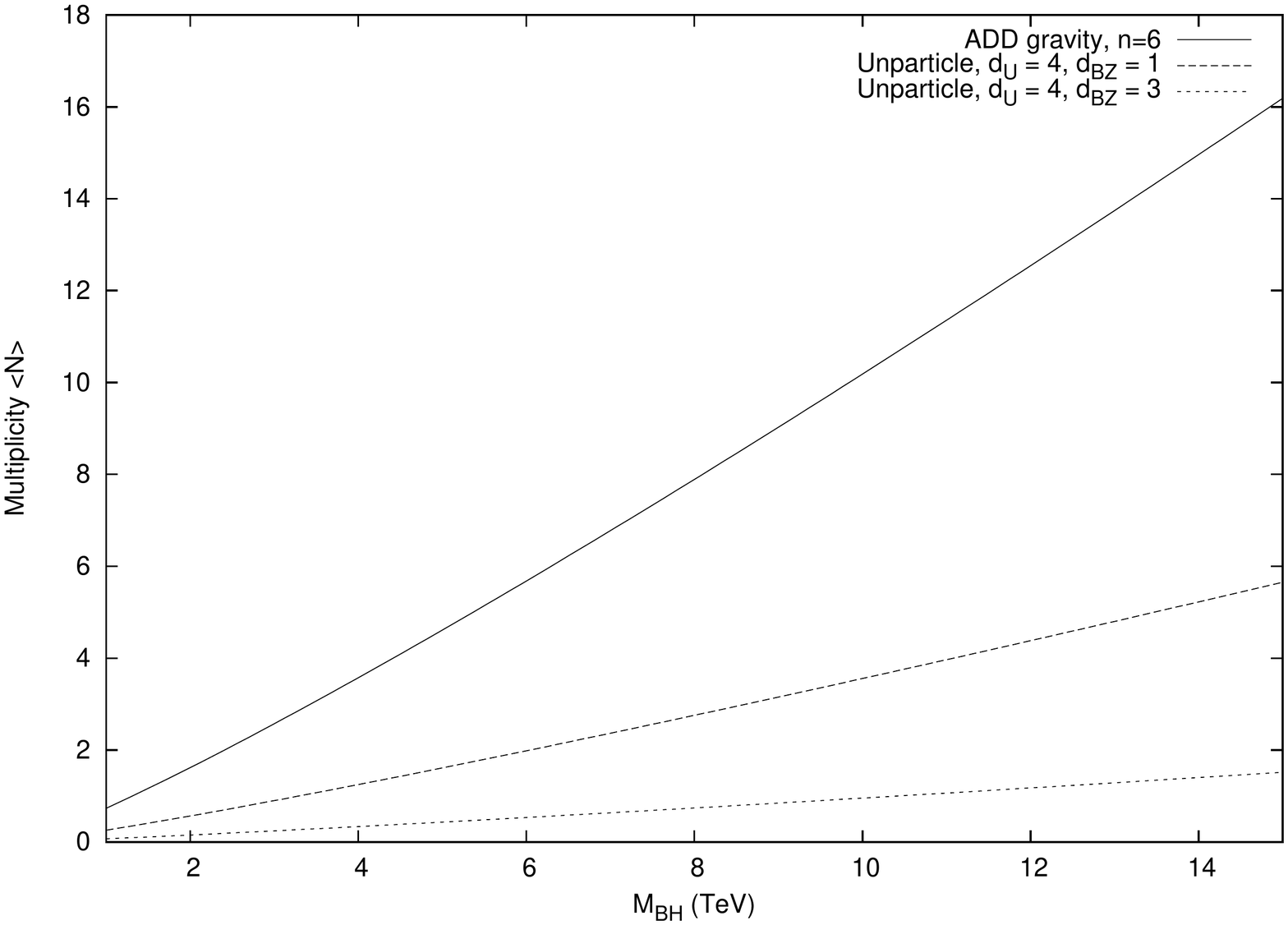}
\caption{Comparison of ADD ($M_D = 1~{\rm TeV}, n=6$) and unparticle-enhanced ($\Lambda_\unp = 1~{\rm TeV}, M_\unp = 10~{\rm TeV}, \du = 4, d_{\rm BZ} = 1,3$) decay multiplicities for LHC black holes for an integrated beam luminosity of $30~{\rm fb}^{-1} {\rm yr}^{-1}$.}
\label{fig3}
\end{center}
\end{figure}

Lastly, it is appropriate to consider the decay modes of the black holes formed by each mechanism.  The multiplicity of particles produced during the evaporation process of an unparticle-enhanced black hole may be compared to that for extra-dimensional black holes.  In the latter case, it has been shown that in the absence of grey-body factors, the multiplicity is well-approximated as \cite{landsberg}
\beq
\langle N \rangle_{\rm ED} \approx \frac{M_{BH}}{2T_{\rm ED}} = \frac{2\sqrt{\pi}}{n+1} \left(\frac{\Mbh}{\Mpl}\right)^\frac{n+2}{n+1} \left[\frac{8\Gamma\left(\frac{n+3}{2}\right)}{n+1}\right]^\frac{1}{n+1}~, 
\label{edmult}
\eeq
Using the definition (\ref{unptemp}) of $T_\unp$, the associated unparticle-driven multiplicity can be written
\beq
\langle N\rangle_\unp = 
\frac{2\pi}{2\du-1} 
 \left(\frac{M_{\rm BH}}{\Lambda_\unp}\right)^\frac{2\du}{2\du-1}  
 \left[2\Gamma_{\du}\left(\frac{M_\unp}{\Lambda_\unp}\right)^{2d_{\rm BZ}} \right]^\frac{1}{2\du-1} 
\label{unpmult}
\eeq
Figure~\ref{fig3} compares the decay multiplicities for black holes of mass $M_{BH}$ in both scenarios, with $n= 6$ and $\du = 4$.   Since the multiplicities are on average a factor of 3 or higher different between the two, this should in principle also provide a method of differentiation.    It must be noted, however, that the $\bra N\ket_\unp < 2$ for a large portion of the parameter space in question violate momentum conservation and are thus unphysical.  These rise above the threshold value only for large black hole masses $\Mbh \geq 8$, and small Banks-Zaks dimension.  In this case, unparticle-driven black hole evaporation will result primarily in pair production of massive particles, making it technically difficult to extract the signal from the background.  Such low multiplicities for even the most liberal choices of parameters places severe constraints on the detectability of unparticle-driven black hole formation, but nevertheless can help in mechanism differentiation.  Future studies that include greybody factors can shed better light on these figures.

Although this presentation explores the idealized case $\Mpl = \Lambda_\unp = 1~$TeV, it may be repeated for arbitrary values of these parameters with similar conclusions.  There are possibly conditions on $\Mpl$ and $\Lambda_\unp$ such that even the intercepts are equal, but this case is tantamount to a new ``fine-tuning'' problem and is unprobable.  Forthcoming data from the LHC can easily be applied to these two cases to make a decisive conclusion on which mechanism, if any, has produced mini black holes.

\section{The future of unparticle physics}
In closing, a few thoughts on the utility of unparticle physics are in order.  The attractiveness of a theory that can re-create extra-dimensional phenomenology in a traditional four-dimensional spacetime is growing in the literature, if for no other reason than we may not see evidence for extra dimensions at the LHC (or anywhere else).  The recently-proposed mechanism of Calumet, Hsu, and Reeb \cite{chr2} postulates a TeV-scale modification to gravity in 4D via interactions of matter with a large number of hidden-sector particles.  The notion of dimensional (de)construction \cite{decons} loosely hints at a similar theme, in which new dimensions are dynamically generated by fundamentally four-dimensional field theories.  This differs with the unparticle-generated ``dimensions'' in that the fundamentally four-dimensional theories appear higher-dimensional at low energies instead of high.  Furthermore, it is not yet clear that the new ``dimensions'' generated by the unparticle couplings follow the formal definitions of the (de)constructed dimensions, beyond fixing the power-law behavior of the potential.  Indeed, it is uncertain how -- if at all -- one might conceptualize the propagation of a particle in a real-valued dimensional spacetime.  

Nevertheless, the importance of unparticle physics must be stressed: it offers a physical mechanism that can induce fractional dimension-behavior in an integral-dimension universe.  This opens the door to a myriad of yet--to-be-explored concepts, as well as provides ground-support for those that have been heretofore mathematical artifacts ({\it e.g.} theories involving ``fractal'' dynamics).  Unparticle physics provides a launching point from which new physics may be discovered well beyond the electroweak scale, in parameter spaces never before accessible in traditional theories.

\pagebreak
\noindent{\bf Acknowledgments}\\
JRM is supported by a Cottrell College Science Award from Research Corporation.

\pagebreak

\begin{thebibliography}{99}
\bibitem{add} N. Arkani-Hamed, S. Dimopoulos, G. Dvali, Phys. Lett. {\bf B 429}, 263 (1998); 
 I. Antoniadis, N. Arkani-Hamed, S. Dimopoulos, G. Dvali, Phys. Lett. {\bf  B 436},  257-263; 
 N. Arkani-Hamed, S. Dimopoulos, G. Dvali, Phys. Rev. {\bf D 59}, 086004 (1999)
\bibitem{fischler} T. Banks and W. Fischler, arXiv:hep-th//9906038 (1999)
\bibitem{landsberg} S.\ Dimopoulos and G. Landsberg, Phys.\ Rev.\ Lett.\ {\bf 87}, 161602 (2001); G. Landsberg, J.\ Phys.\ {\bf G32} R337-R365 (2006)
\bibitem{giddings} S.\ B.\ Giddings and S.\ Thomas, 	Phys. Rev. {\bf D65}, 056010 (2002)
\bibitem{georgi} H. Georgi, Phys.\ Rev.\ Lett.\ {\bf 98}, 221601 (2007); Phys.\ Lett.\ {\bf B 650}, 275--278 (2007)
\bibitem{cheung1}  K.~Cheung, W.~Y.~Keung and T.~C.~Yuan,  Phys.\ Rev.\ Lett.\  {\bf 99}, 051803 (2007)
\bibitem{bz}   T.~Banks and A.~Zaks,  Nucl.\ Phys.\  B {\bf 196}, 189 (1982)
\bibitem{kraz}   N.~V.~Krasnikov,  Int.\ J.\ Mod.\ Phys.\  A {\bf 22}, 5117 (2007)
\bibitem{nikolic}  H.~Nikolic, Mod.\ Phys.\ Lett.\ {\bf A 23 (31)}, 2645-2649 (2008), arXiv:0801.4471 [hep-ph]
\bibitem{mcdonald2} J. McDonald, arXiv:0805.1888 [hep-ph] (2008)
\bibitem{georgi2}  H.~Georgi and Y.~Kats,  Phys.\ Rev.\ Lett.\ {\bf 101}, 131603 (2008); arXiv:0805.3953 [hep-ph]
\bibitem{goldberg}  H.~Goldberg and P.~Nath,  Phys.\ Rev.\ Lett.\  {\bf 100}, 031803 (2008)
\bibitem{hsu1}  N.~G.~Deshpande, S.~D.~H.~Hsu and J.~Jiang,  Phys.\ Lett.\  B {\bf 659}, 888 (2008)
\bibitem{damora}  S.~Das, S.~Mohanty and K.~Rao,  Phys.\ Rev.\  D {\bf 77}, 076001 (2008)
\bibitem{jrmplb}   J.~R.~Mureika,  Phys.\ Lett.\  B {\bf 660}, 561 (2008)
\bibitem{cheung2}  K.~Cheung, W.~Y.~Keung and T.~C.~Yuan,  Phys.\ Rev.\  D {\bf 76}, 055003 (2007)
\bibitem{rindler} W.~Rindler, {\it Relativity: Special, General, and Cosmological}, Oxford University Press, New York (2001)
\bibitem{nakayama}  Y.~Nakayama,  Phys.\ Rev.\  D {\bf 76}, 105009 (2007)
\bibitem{grinstein} B.~Grinstein, K.~Intriligator and I.~Z.~Rothstein,  Phys.\ Lett.\  B {\bf 662}, 367 (2008)
\bibitem{mcdonald1} J.~McDonald, arXiv:0709.2350 [hep-ph] (2008)
\bibitem{chr1}    A. Chamblin, S. W. Hawking and H. S. Reall, Phys. Rev. D 61, 065007 (2000); 
\bibitem{rsbh}    L. A. Anchordoqui,H. Goldberg, A. D. Shapere, Phys. Rev.{\bf D66}, 024033 (2002) 
\bibitem{myersperry}  R.~C.~Myers and M.~J.~Perry,  Annals Phys.\  {\bf 172}, 304 (1986).
\bibitem{poisson} E.\ Poisson, {\it A Relativist's Toolkit: The Mathematics of Black Hole Mechanics}, Cambridge University Press, New York (2008)
\bibitem{mannkerner}  R.~Kerner and R.~B.~Mann,  Class.\ Quant.\ Grav.\  {\bf 25}, 095014 (2008)
\bibitem{chr2} X.\ Calumet,  S. D. H. Hsu, d. Reeb, Phys.\ Rev.\ {\bf D 77}, 125015 (2008)
\bibitem{decons} N. Arkani-Hamed, A.G. Cohen and H. Georgi, Phys. Rev. Lett. {\bf 86} (2001) 
\end{thebibliography}
 \end{document}